# Light-induced Magnetic Phase Transition in van der Waals Antiferromagnets

Jiabin Chen[1,2,3], Yang Li[3], Hongyu Yu[1,2], Yali Yang[1,2], Heng Jin[3], Bing Huang[3,4]*, Hongjun Xiang[1,2,5]*

[1]*Key Laboratory of Computational Physical Sciences (Ministry of Education), Institute of Computational Physical Sciences, State Key Laboratory of Surface Physics, and Department of Physics, Fudan University, Shanghai 200433, China*

[2]*Shanghai Qi Zhi Institute, Shanghai 200030, China*

[3]*Beijing Computational Science Research Center, Beijing 100193, China*

[4]*Department of Physics, Beijing Normal University, Beijing 100193, China*

[5]*Collaborative Innovation Center of Advanced Microstructures, Nanjing 210093, China*

*E-mail: bing.huang@csrc.ac.cn, hxiang@fudan.edu.cn

**Abstract**

Based on a simple tight-binding model, we propose a general theory of light-induced magnetic phase transition (MPT) in antiferromagnets based on the general conclusion that the bandgap of antiferromagnetic (AFM) phase is usually larger than that of ferromagnetic (FM) one in a given system. Light-induced electronic excitation prefers to stabilize the FM state over the AFM one, and once the critical photocarrier concentration ($\alpha_c$) is reached, an MPT from AFM phase to FM phase takes place. This theory has been confirmed by performing first-principles calculations on a series of two-dimensional (2D) van der Waals (vdW) antiferromagnets and a linear relationship between $\alpha_c$ and the intrinsic material parameters is obtained. Importantly, our conclusion is still valid even considering the strong exciton effects during photoexcitation. Our general theory provides new ideas to realize reversible read-write operations for future memory devices.

**Main text**

Discovering long-range magnetism in two-dimensional (2D) materials is one of the most significant developments in the field of magnetism in recent years [1,2]. Once the magnetic phase transition (MPT) in 2D magnets can be achieved, they can be widely applied to design nanoscale spintronic and memory devices. In principle, because of their 2D limit thickness, their MPT can be effectively controlled or switched by many external perturbations [3]. In practice, the engineering of MPT in different 2D magnets using various approaches, e.g., external electric [4], magnetic [5], and strain fields [6,7], is developing rapidly. However, these methods are either applicable for specific material systems or are not easy to control precisely in practice. Thus, developing easily accessible approaches for engineering MPT in 2D magnets remains highly desirable.

Illumination is particularly efficient in reversibly manipulating the physical properties of materials. For example, illumination can control topological phase transitions [8], charge transport behaviors [9], and charge order transitions [10,11]. For a long time, the magneto-optical effect has been regarded as a powerful technique for detecting the magnetic properties in solids [12]. Recently, illumination was demonstrated as a feasible way with remoteness, rapidity, and lower energetic consumption in controlling the magnetic properties of layered vdW magnets [13,14]. For example, light can attenuate the rhombohedral lattice distortion by rapidly weakening the magnetic order in NiO [15], and the light-induced ferromagnetism in moiré superlattices has been observed [16]. In addition, light-induced demagnetization in ferromagnetic (FM) materials has also been proposed [17]. The realization of light-induced MPT in 2D vdW systems is still in its early stage. In particular, a general theory for realizing MPT between the two most fundamental magnetic phases, FM and AFM orders, is still lacking.

In this Letter, we first employ a simple tight-binding model to discover a general mechanism of the photoexcitation-induced AFM to FM phase transition. This mechanism is closely related to the fact that the bandgap of the AFM configuration is generally larger than that of the FM configuration. Using density functional theory (DFT) to obtain magnetic properties of the ground state and constrained density functional theory (cDFT, as described in ref. [18]) to mimic the thermalized photocarrier population, we propose that vdW layered antiferromagnets are the ideal platforms to study magnetic phase transition under laser illumination. We demonstrate the phase transitions from AFM state to FM state in monolayer MnI$_2$ and MnP$X_3$($X$=S, Se, Te) and bilayer CrI$_3$, MnBi$_2$Te$_4$ when the photocarrier concentration $n_e$ reaches the order of $10^{12} e^-/cm^2$. Note that photocarriers with the concentration of (2-3) × $10^{12} e^-/cm^2$ have been experimentally achieved in the case of 2D MoS$_2$ [19]. More calculations details can be found in the supplemental materials (SM).

**General theory of light-induced MPT.** As shown in Fig. 1, to identify how the bandgap (G$_{diff}$) difference and total energy difference (E$_{diff}$) between AFM and FM configurations determine the ground state of the system, we introduce a two-site tight-binding model [20] to simulate the simplest magnetic semiconductor system. The mean-field Hamiltonians for the AFM and FM states are given as

$$H = -t \sum_{\sigma=\uparrow,\downarrow} (a^+_{1\sigma} a_{2\sigma} + h.c.) + U/2 \sum_{i=1,2} S_i(a^+_{i\downarrow} a_{i\downarrow} - a^+_{i\uparrow} a_{i\uparrow}) \quad (1)$$

where $i$ represents the two neighboring magnetic sites, $U$ represents the on-site repulsive energy and $t$ represents the hopping integral, $S_i$ (+1 or -1) denotes the spin direction of the $i$-th site. For the FM state, $S_1$=$S_2$=1, while $S_1 = 1$ and $S_2 = -1$ for the AFM state. By diagonalizing the Hamiltonian and considering the usual condition $U \gg t$ in semiconductors and insulators, we can get

$$E_{AFM} = -\sqrt{U^2 + 4t^2} \approx -U - \frac{2t^2}{U}, \quad E_{FM} = -U \quad (2)$$

$$G_{AFM} = \sqrt{U^2 + 4t^2} \approx U + \frac{2t^2}{U}, \quad G_{FM} = U - 2t \quad (3)$$

where E$_{AFM}$ and E$_{FM}$ represent the total energy in AFM and FM configurations respectively, and G$_{AFM}$ and G$_{FM}$ represent the bandgap in AFM and FM configurations respectively. Consequently, we can obtain the energy difference (E$_{diff}$) and bandgap difference (G$_{diff}$) between the AFM and FM states,

$$E_{diff} = E_{AFM} - E_{FM} = -\frac{2t^2}{U} \quad (4)$$

$$G_{diff} = G_{AFM} - G_{FM} = \frac{2t^2}{U} + 2t \tag{5}$$

As indicated in Eq. (4), we can see that the energy of the AFM state is lower than that of the FM state, which is consistent with the fact that the kinetic exchange usually leads to antiferromagnetism. Interestingly, Eq. (5) indicates that the bandgap of the FM state is smaller than that of the AFM state, in agreement with the fact that the FM state is usually more metallic [e.g., in colossal magnetoresistance (CMR) manganese oxides].

We now assume that α number of electrons is photoexcited from the valence bands (VB) to the conduction bands (CB) under light illumination, then the new energy difference ($E'_{diff}$) between the AFM and FM states becomes

$$E'_{diff} = E_{diff} + \alpha G_{diff} = -\frac{2t^2}{U} + \alpha\left(\frac{2t^2}{U} + 2t\right). \tag{6}$$

The critical condition is $E'_{diff} = 0$, i.e., $\alpha_c = -\frac{E_{diff}}{G_{diff}}$, which indicates that once $\alpha \geq \alpha_c$ electron is excited, and the magnetic phase transition from the AFM state to the FM state can be realized. Here we assume that the band structure change induced by photoexcitation is negligible which is reasonable as long as the photocarrier concentration is not large ($n_e \ll 1e^-$/f.u.). It is worth noting that the thermodynamic process rather than the dynamic process is considered here, i.e., the time-dependent electron spin transfer and flipping process are not included. This is because the system can always reach its lowest-energy magnetic state after a sufficiently long relaxation time. Under illumination, the electrons will be excited from VB to CB in the order of 1 ns even in indirect bandgap semiconductors, and the excited high-energy electrons and holes will fall back to CB minimum (CBM) and VB maximum (VBM) in a shorter time (~1ps), respectively. Furthermore, the timescale required for the photocarrier recombination (1ns − 1μs) is much longer than that of excitation [21]. Given this fact, here we focus on the steady state with constant hole and electron concentration at VBM and CBM under a persistent light illumination, without investigating the ultrafast response on picosecond time scale.

**Light-induced MPT in monolayer MnI$_2$.** Monolayer MnI$_2$ with the P-3m1

symmetry was reported to exhibit a screw-type AFM ground state. This magnetic structure induces ferroelectric polarization along the [110] direction of MnI$_2$ [22–24], which was explained by the general theory proposed for describing the ferroelectric polarization induced by a spin-spiral order [25]. Here, we propose that illumination can be adopted to realize the MPT in monolayer MnI$_2$. As shown in Fig. 2(a), a simple stripe-type collinear configuration is adopted for the AFM state in a $4 \times 4 \times 1$ supercell. More computational details can be found in supplemental materials (SM). As shown in Fig. 2(b), the CBM and VBM of both AFM and FM states are mainly contributed by the cation Mn *d* and anion I *p* orbitals, respectively. Therefore, the photoexcited electron is transferred between anion and cation. This is slightly different from the scenario described by the TB model [see Eq. (1)] where only cation orbitals are considered. However, it does not influence our main conclusion on MPT because it is based on the fact that the bandgap of AFM configuration is larger than that of FM configuration. Our DFT calculations show that $E_{\text{diff}}$ and $G_{\text{diff}}$ are $-0.077$ eV and $0.321$ eV base on Fig. 1(a) configurations, respectively. According to the model we proposed [see Eq. (6)], the critical photocarrier for MPT is $\alpha_c = 0.239 \ e^-$, which is equivalent to $5.015 \times 10^{12} e^-/cm^2$. When $\alpha \geq \alpha_c$, the total energy of the FM state will be lower than that of the AFM state. We confirm this result by performing the cDFT calculations where the presence of electrons and holes is explicitly taken into account. Our cDFT calculations indeed show that the total energy difference $E'_{\text{diff}}$ is zero when we set $\alpha_c = 0.239 e^-$ for both FM state and AFM state [see the star point in Fig. 2(c)]. Besides, we verified that the light-induced MPT between the non-colliear AFM state and FM state will also occur by considering a helical spin-spiral AFM state with q = (1/3,0,0) and q = (1/3,1/3,0)[25] (see SM).

It is well-known that the exciton effect in 2D vdW materials is strong due to the reduced screen effect of electron-hole interaction. Therefore, it is important to further verify whether our above conclusion is still valid under the consideration of the exciton effect. Accordingly, we perform a series of first-principles GW and Bethe-Salpeter equation (BSE) [26–28] calculations for MnI$_2$ monolayer. The calculated bandgaps of AFM (FM) configuration under DFT, GW, GW-BSE methods are 2.23 (1.91) eV, 4.28 (4.08) eV, and 3.17 (3.06) eV, respectively. Note that the value of 3.06 eV is estimated through the scaling law between bandgap and exciton binding energy in 2D semiconductors [29], since the CBM and VBM in the FM configuration belong to different spin channels and our calculations cannot include the electron hole entanglement between different spins. We emphasize that the calculated results of AFM configuration comply with this law very well. As shown in Fig. 2(c), we plot the energy difference between the AFM state and FM state $E'_{\text{diff}}$ as a function of the number ($\alpha$) of the photoexcited electrons. We can see that the FM state will become more stable if enough photocarrier is present even when the exciton effect is included. The linear slope of DFT, GW, GW-BSE curves decreases, which indicates that the electron-hole correlation effect slightly increases the $\alpha_c$ required for MPT.

**Light-induced MPT in other vdW antiferromagnets.** In addition to monolayer MnI$_2$, we will demonstrate that the idea to realize AFM-FM transition using light is general. To further illustrate this, we perform calculations for different 2D antiferromagnets, whose $-\frac{E_{\text{diff}}}{G_{\text{diff}}}$ is plotted as a function of $\alpha_c$ in Fig. 3 [see details in SM]. Remarkably, we obtain a relational line with slope 1 and intercept 0. These systems vary widely in element types, atomic bond lengths and crystal structures, indicating the generality of our proposed theory. Among these materials, monolayer MnP$X_3$($X$=Se, S, Te) exhibits Néel antiferromagnetism [30] and valley-dependent optical properties [31]. MnPTe$_3$ was reported to undergo MPT by varying the electric field [32], here we predict that light illumination is also a very promising approach.

In addition to the monolayer magnetic materials, multilayered materials with interlayer AFM coupling were also reported to display exotic phenomena. For example, giant nonreciprocal second-harmonic generation has been observed in AFM bilayer CrI$_3$ [33], and modulation of interlayer antiferromagnetism can lead to drastic changes of excitonic transitions in CrSBr bilayers [34]. Therefore, besides the light-induced intralayer AFM-FM transition, we also perform calculations for several AFM multilayers, including the CrI$_3$ bilayer and MnBi$_2$Te$_4$. We find that light illumination can also change their interlayer AFM coupling to FM coupling. Among them, CrI$_3$ presents to be an intriguing case. Although bulk CrI$_3$ is FM, the interlayer coupling becomes AFM when the system is thinned down to a few atomic layers [35], and its interlayer exchange can be tuned by stacking [36–38]. Here, we only consider the high temperature phase with the AFM interlayer coupling. Our calculations show only a photoelectron concentration of $\alpha_c = 1.073 \times 10^{-2} e^-/\text{f.u.}$ (equivalent to $n_e = 2.523 \times 10^{12} e^-/cm^2$) is required for the MPT in bilayer CrI$_3$. As CrI$_3$ bilayer can realize the appearance and disappearance of net magnetization through the modulation of light, it is a promising candidate material for next generation photodetectors. Besides, MnBi$_2$Te$_4$ was reported to be topological axion state and antiferromagnetism in even number of layers films while the QAH effect and ferromagnetism in odd number of layers films [39–44]. We also find that the MPT of MnBi$_2$Te$_4$ can be achieved under illumination (see Fig. 3), similar to the CrI$_3$ bilayer case.

**Discussion**. We note that light can induce MPT in materials as long as the bandgap of the material with ground-state magnetic configuration is larger than that of a higher energy magnetic state. This mechanism is independent of the dimension of materials. For example, 3D CuFeO$_2$ is reported to be a geometrically frustrated triangular lattice antiferromagnet [45]. In the presence of enough photocarrier, it will become FM [see SM]. In addition, the change of magnetism caused by light will indirectly affect other fundamental properties, such as ferroelectricity and topologic phase. The AFM order in MnI$_2$ breaks the spatial inversion symmetry and induces ferroelectricity, therefore, the change of magnetic phase under light allows for tuning its ferroelectricity. The AFM-FM phase transition of MnBi$_2$Te$_4$ system was suggested to induce the transition from axion insulator to Chern insulator phase[39], thus light can be an excellent tool to realize this transformation.

To summarize, we present a general theory of MPT in antiferromagnets. Importantly, we derive a simple linear relationship between critical excited photocarrier concentration $\alpha_c$ and the intrinsic material parameters (including energy difference and bandgap difference between AFM and FM states). This general relationship is well confirmed in a series of first-principles calculations for 2D antiferromagnets, regardless of the types and layers of materials, which can provide a universal guiding basis for future optoelectronic device designs such as photodetectors, magnetic storage devices, etc.

**ACKNOWLEDGMENTS**

Work at Fudan is supported by NSFC (Grants No. 11825403, No. 11991061, and No. 12188101) and Guangdong Major Project of Basic and Applied Basic Research (Future functional materials under extreme conditions - 2021B0301030005). We also acknowledge the support from the NSAF (Grant No. U1930402).

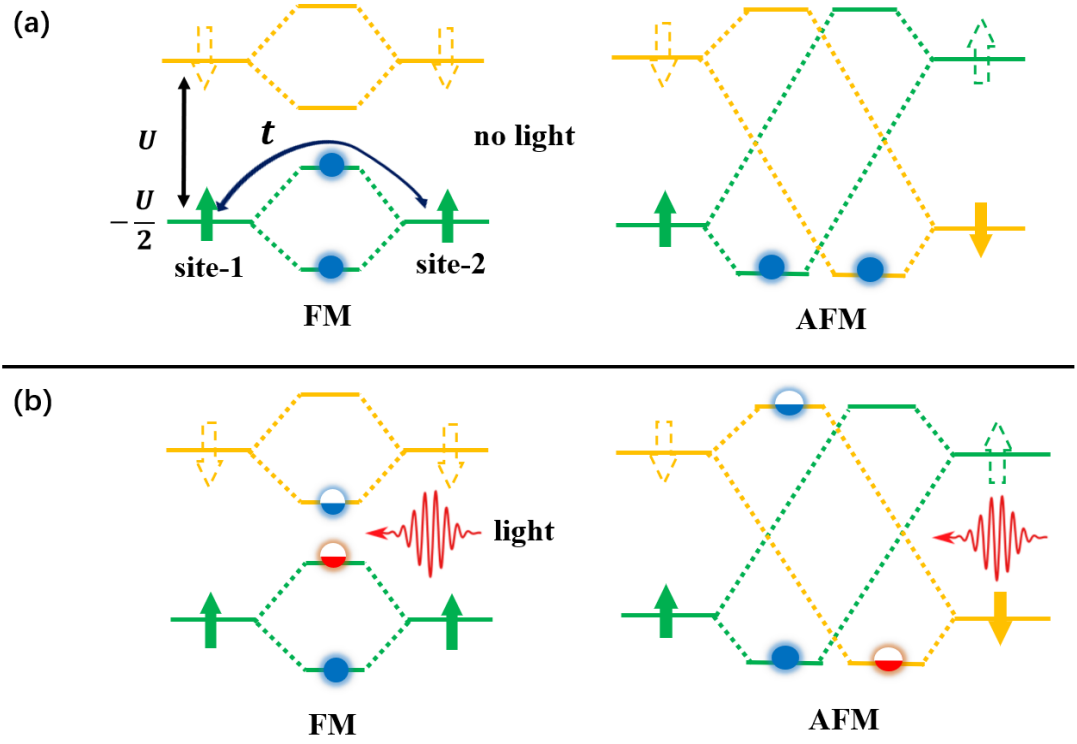

FIG. 1. Schematic diagrams of orbital evolution in a double-site model. (a) FM configuration with no light (left) and AFM configuration with no light (right). (b) FM configuration with light (left) and AFM configuration with light(right). A partial electrons were excited. Green and orange lines represent spin-up and spin-down orbitals, respectively. Blue and red dots represent electrons and holes, respectively.

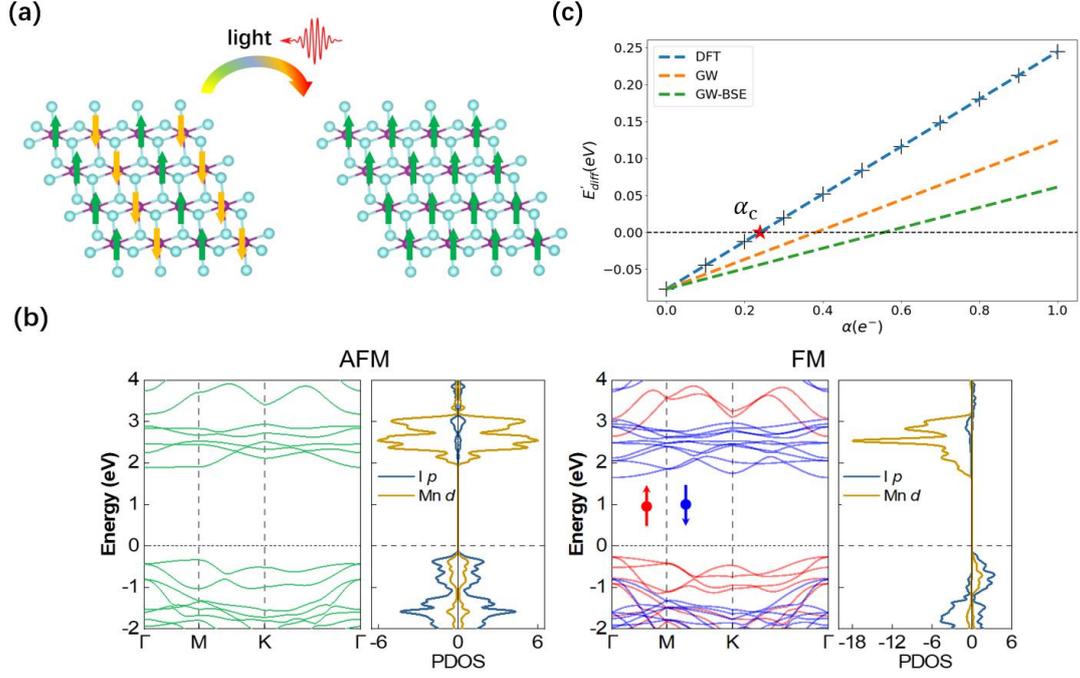

FIG. 2. (a) Crystal structure of (left) AFM and (right) FM configurations of monolayer MnI$_2$ from the top view. Purple and blue balls represent Mn and I atoms, respectively. Green and yellow arrows represent the spin up and spin down, respectively (b) DFT-calculated band structure and density of states of monolayer MnI$_2$ under (left) AFM and (right) FM configurations. (c) Dependencies of the MnI$_2$ energy difference between the AFM state and the FM state of DFT, GW, and GW-BSE as a function of the excited electron concentration. The cross marked points are cDFT calculation results, and the dotted line is derived from the bandgap of DFT and model mentioned above. The star point represents the number of photoexcited electrons with $E'_{diff} = 0$, which are based on MnI$_2$ 4×4 supercell.

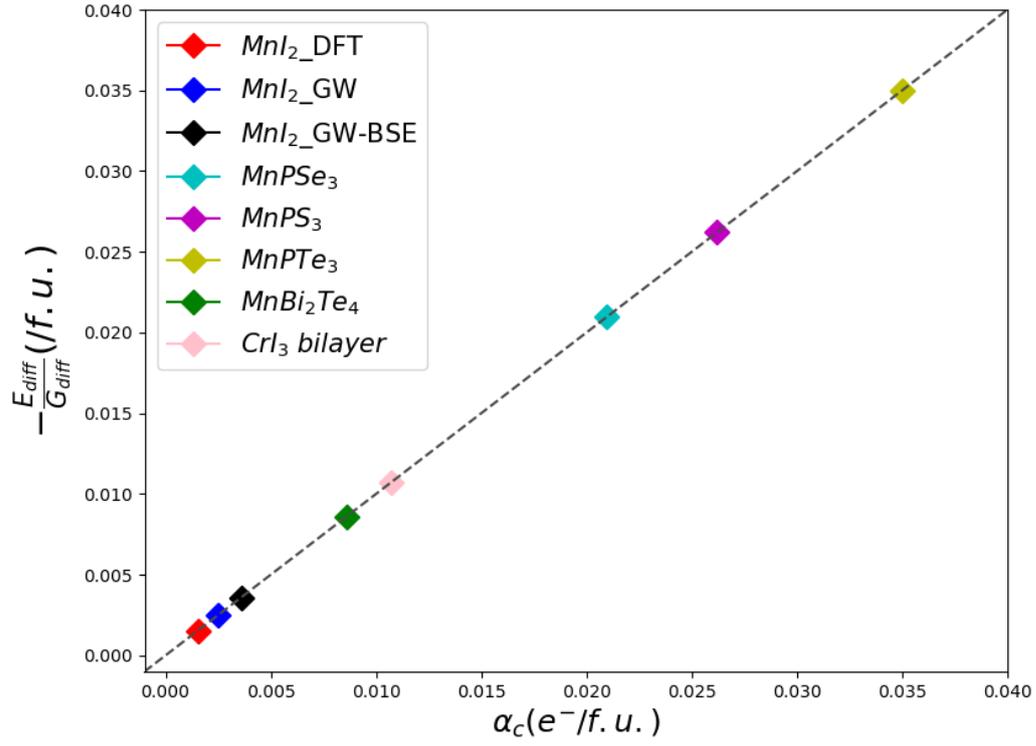

FIG. 3. Linear relationship between excited electrons number $\alpha_c$ and ratio of ground state energy difference to bandgap difference $-E_{diff}/G_{diff}$. Calculations cover monolayer vdW materials $MnI_2$, $MnPX_3$ ($X$=Se, S, Te), and multilayer materials $MnBi_2Te_4$ and $CrI_3$ bilayer of HT phase.